\documentclass[prl,aps,twocolumn]{revtex4}
\usepackage[pdftex]{graphicx}
\usepackage{amssymb}

\def \aap      {A\&A }

\def \aj       {AJ}
\def \apj      {ApJ }
\def \apjl     {ApJL }
\def \apjs     {ApJS }

\def \mnras    {MNRAS }

\def \prd      {Phys. Rev. D }

\def\aapr{A\&A Rev.}
\def \pasa     {PASA}

\def \gr{$\gamma$-ray}

\date{}

\begin{document} 
\title{Supernova origin of cosmic rays from a $\gamma$-ray signal in the Constellation III region of the Large Magellanic Cloud} 
\author{Andrii Neronov}
\affiliation{Astronomy Department, University of Geneva, Ch. d'Ecogia 16, Versoix, Switzerland}
\begin{abstract}
 Cosmic rays could be produced via shock acceleration powered by supernovae. The supernova hypothesis implies that each supernova injects on average some $10^{50}$ erg in cosmic rays, while the shock acceleration model predicts a powerlaw cosmic ray spectrum with the slope close to 2.  Verification of these predictions requires measurements of spectrum and power of cosmic ray injection from supernova population(s). Here we obtain such measurements based on \gr\   observation of Constellation III region of Large Magellanic Cloud. We show that \gr\ emission from this young star formation region originates from cosmic rays injected by approximately two thousand supernovae, rather than by massive star wind powered superbubble pre-dating supernova activity.  Cosmic ray  injection power is found to be $(1.1_{-0.2}^{+0.5})\times 10^{50}$~erg/supernova (for the estimated interstellar medium density 0.3 cm$^{-3}$). The spectrum is a powerlaw with slope $2.09_{-0.07}^{+0.06}$. This agrees with  the model of particle acceleration at supernova shocks and provides a direct proof of the supernova origin of cosmic rays.  
\end{abstract}

\date{}
\maketitle 

The bulk of the flux of cosmic rays reaching the Earth is believed to be generated by shock acceleration process \cite{fermi49,krymskii,bell1} operating in supernova powered sources \cite{baade_zwicky,ginzburg,blasi} or and/or superbubbles of star formation \cite{bykov}. However, direct verification of  either supernova or superbubble scenario is difficult because information on sources is erased during propagation of cosmic rays through the interstellar medium toward the Earth \cite{berezinskii}. 

It is also difficult to derive the spectrum and power of cosmic ray injection from \gr\ signal of astronomical sources like supernova remnants because of uncertainties of individual source parameters (e.g. distance, density of the ambient interstellar medium and/or of the pre-existing stellar wind bubble structure) and distortion of the spectrum by the effect of escape from the source  \cite{snr1,tavani}. Enhanced \gr\ emission from interactions of hadronic cosmic rays is often observed at locations of molecular clouds located near or interacting with supernova remnants. In this case the \gr\ flux depends on the uncertain density of the cloud and on the uncertain details of propagation of cosmic rays toward the cloud. Another source of uncertainty is in the separation between the \gr s produced in interactions of high-energy protons via production and decays of pions from and those produced by electrons though the inverse Compton and/or Bremsstrahlung.    	

Cosmic rays injected  into interstellar medium retain their injection spectrum as long as they form a finite size expanding cocoon around the source. Observations of \gr\ flux from cosmic ray interactions with interstellar medium inside such a cocoon potentially provide a possibility of calorimetric measurement of  the spectrum and power of cosmic ray injection from the source \cite{aharonian,tev_cr}. 

However, interpretation of \gr\ data on such cosmic ray cocoons in the Milky Way (like that in the Cygnus region \cite{cygnus}),  is complicated because the \gr\ signal  is  superimposed onto background diffuse \gr\ emission from the Milky Way disk \cite{fermi_diffuse}. Projection effects also superimpose stellar population of different ages and at different distances \cite{orion_arm}  so that it is difficult to separate the \gr\ flux component generated by  youngest cosmic ray sources from that of the older ones.  For example, in the particular case of Cygnus cosmic ray cocoon, the cosmic rays might originate either from a young Cyg OB2 association which has not yet produced supernovae or by one or many supernovae (like, e.g. gamma Cygni), depending on the (uncertain) distance to the supernova(e). 

Clear discrimination between the  cosmic ray injection by acceleration processes operating in stellar-wind powered superbubbles and by supernovae would be possible if young star forming regions would be observed as isolated source on the sky, not superimposed onto other potential sources. Measurement of the specta of \gr\ emission from such isolated regions might even provide a timing of the moment of the onset of cosmic ray injection with several Myr precision. This is possible because the spectrum of cosmic rays residing in the region is modified by propagation effects only after some time delay, estimated as the time needed for cosmic rays to diffuse out of the production region. 

The projection problem is removed and young star forming regions could be observed as isolated sources in Large Magellanic Cloud (LMC), which is the nearest galaxy with on-going star formation and with the disk observed almost face-on  \cite{fermi_lmc1,fermi_lmc2}. Contrary to the Milky Way, absense of projection effects allows to better control the details of the star formation history at the sites of on-going star formation in the  LMC \cite{harris08,harris09}.  

In what follows we use \gr\ data of  Fermi Large Area Telescope (LAT) \cite{atwood} to obtain a calorimetric measurement of the spectrum and overall energy injected in cosmic rays by a population of supernovae which have exploded during the last several million years in one of the young star forming regions in the LMC, the Constellation III (Con III) region. We show that this measurement provides a full test of the supernova scenario of the origin of cosmic rays first proposed in 1934 by Baade \& Zwicky \cite{baade_zwicky}. 

\begin{figure}
\includegraphics[width=\linewidth]{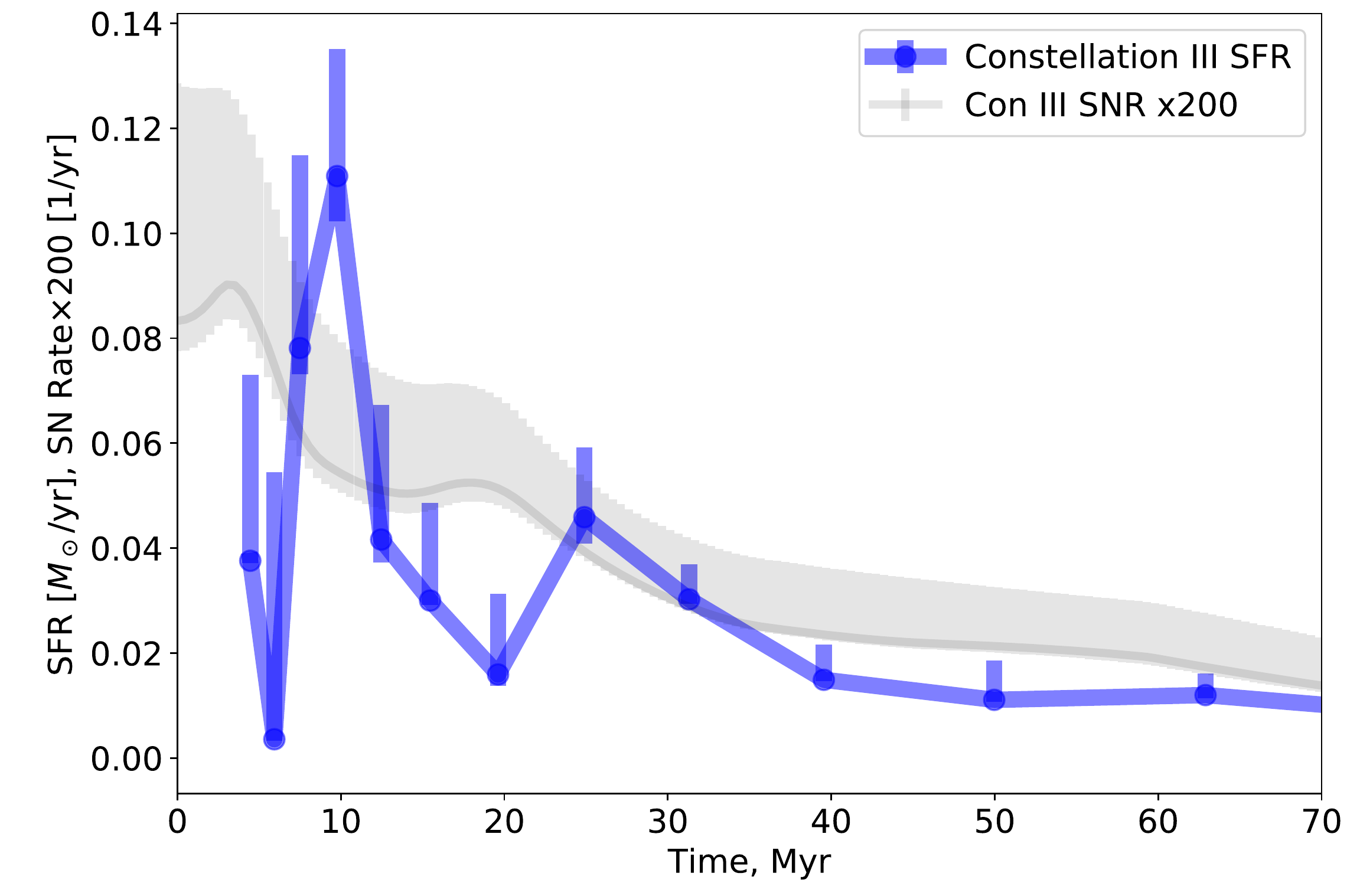}
\caption{History of the SFR of the Con III region from Ref. \cite{harris08}. Grey shaded band shows the supernova rate time evolution derived from the SFR. Band width indicates the uncertainty of the supernova rate estimate. }
\label{fig:sn_rate}
\end{figure}

Con III is one of the youngest star forming regions in LMC.  Its star formation rate (SFR) history  \cite{harris08,harris09} is shown in Fig. \ref{fig:sn_rate}. {Refs. \cite{harris08,harris09}  have derived  SFR based on analysis of color-magnitude diagrams of stellar complexes in Con III region.} The peak of the SFR which occurred ${\cal T}_{SFR}\simeq 10$~Myr ago leads to an increase of supernova rate with a time delay 5-10 Myr \cite{leitherr,svensmark} {at ${\cal T}_{SN}\simeq 5$~Myr.} Fig. \ref{fig:sn_rate} shows this increase by the grey curve calculated using the method of Ref. \cite{svensmark}, via convolution of the SFR history with a kernel function describing evolution of supernova rate following an instanteneous star burst.   Con III region is spatially coincident with the hard spectrum "E2"  \gr\ source reported in the Ref.  \cite{fermi_lmc2}.  The \gr\ emission is naturally explained by interactions of cosmic rays with the interstellar medium. Hardness of the source spectrum indicates that the cosmic rays are still contained in a finite size cocoon, like that in the Cygnus region \cite{aharonian,tev_cr,cygnus}. 

Identification of the \gr\ source with Con III allows to refine the analysis and draw important conclusions from the \gr\ measurements. 
For our analysis we have used the data of Fermi/LAT telescope collected between August 4, 2008 and June 1, 2017. The data were processed in the standard way using Fermi Science Tools version v10r0p5 \footnote{https://fermi.gsfc.nasa.gov/ssc/data/analysis/}. Gamma-ray events belonging to the SOURCE class were selected. 

\begin{figure}
\includegraphics[width=\linewidth]{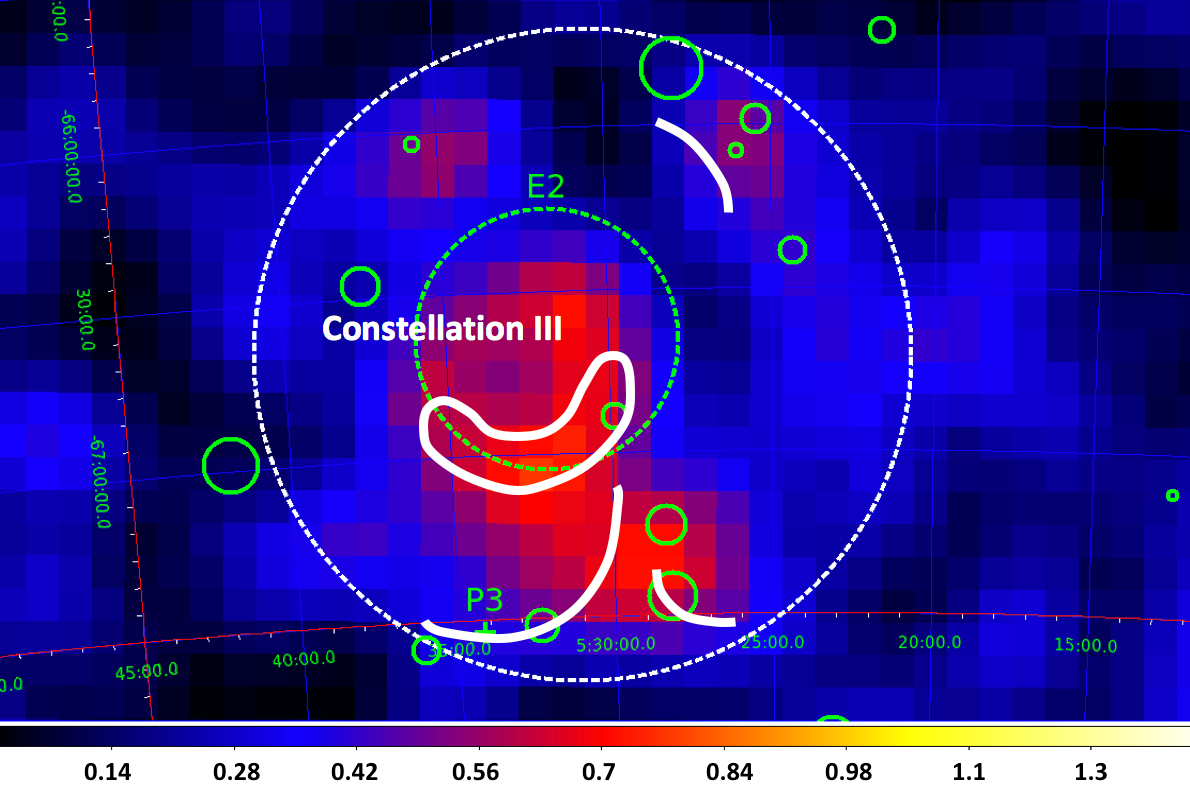}
\caption{LAT countmap of LMC in the energy band  $E>10$~GeV smoothed with 0.3 degree Gaussian.  Green small circles show positions of known supernova remnants  \cite{snr_cat}. White {dashed circle of the radius $1^\circ$} outlines the extent of the Con III region. White arc-like curves show positions of Con III star forming complexes from Ref. \cite{harris08}.  Green dashed circle shows "E2" source \cite{fermi_lmc2}.}
\label{fig:image}
\end{figure}

Fig. \ref{fig:image} shows the \gr\ countmap in the energy range $E>10$~GeV. White circle of the radius $1^\circ$ centered at RA=$82.7^\circ$, DEC=$-66.7^\circ$ shows the extent of the  Con III region for which the star formation history was derived in Refs. \cite{harris08,harris09}.  The \gr\ emission does not exhibit excess at the positions of known supernova remnants listed in Ref. \cite{snr_cat}. Stellar complexes in Con III include several arc-like structures \cite{harris08}  also shown in Fig. \ref{fig:image}.  The morphology of the diffuse \gr\ emission does not repeat that of the stellar arcs. It extends around the arcs. A consistent interpretation of such morphology is in its origin from interactions of cosmic rays spreading into interstellar medium.

\begin{figure}
\includegraphics[width=\linewidth]{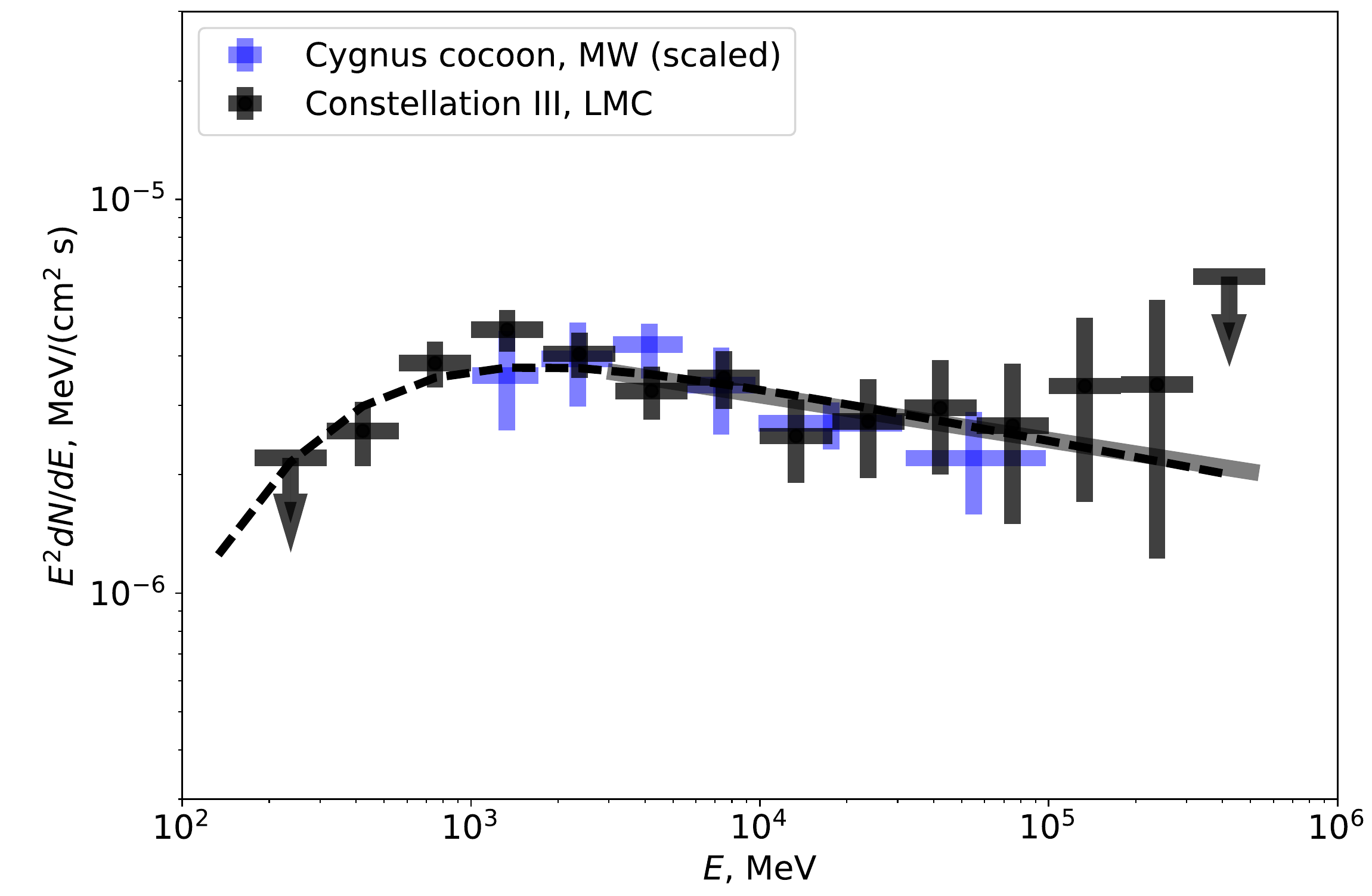}
\caption{{$\gamma$-ray} spectrum of the Con III region (black data points).  Blue semi-transparent data points show rescaled spectrum of Cygnus cocoon \cite{cygnus}. Grey solid line shows the best fit powerlaw spectrum. Dashed curve shows the best fit pion decay spectrum.}
\label{fig:spectrum}
\end{figure}

The source spectrum shown in Fig. \ref{fig:spectrum} was calculated using the unbinned likelihood analysis. The sky model fitted to the data included a set of point sources from the four-year Fermi LAT catalog \cite{fermi_cat} as well as point and extended sources in the LMC found in Ref.  \cite{fermi_lmc2}, except for the LMC disk and the source E2 which is situated at the position of Con III.  For these two sources we have chosen the spatial model which is adapted to the analysis of Ref. \cite{harris08,harris09}. This model takes int account that the LMC disk is observed almost face-on \cite{lmc_disk}. A kiloparsec-wide region of the disk occupied by the Con III  presumably  extends over the entire disk thickness. This suggests spatial models of the two extended sources, large scale LMC disk and Con III region  arranged like a  puzzle game. The Con III region circle  fits into a matching hole in the larger disk of the LMC. The Con III disk has radius $1^\circ$ and is centered at RA=$82.7^\circ$, DEC=$-66.7^\circ$, the centre of Con III region defined in Ref. \cite{harris08}. The LMC disk has radius $4^\circ$ and is centered at the position of the Fermi LMC disk source and extending over the size of the HI disk of the LMC \cite{hi}.  In the absence of a-priori knowledge of the surface brightness distribution of the two extended sources, we have chosen a simple flat radial brightness profiles for both the Con III and the LMC disks. 

A point source P3 reported in Ref. \cite{fermi_lmc2} is situated at the border of the Con III region (see Fig. \ref{fig:image}, within one of the stellar arcs. It is not clear if this soft spectrum point source is a part of the diffuse emission or it is indeed an isolated bright source (e.g. a pulsar wind nebula). We include this source as a point source in the likelihood analysis.  

The shape of the spectrum of Con III source agrees with the previous measurement of the E2 source spectrum  \cite{fermi_lmc2} but the measurement extends to higher energies due to longer exposure and larger signal collection region.  In the energy range above 2~GeV the spectrum is well fit by a  powerlaw  $dN/dE=A (E/1\mbox{ GeV})^{-\Gamma}$ with the slope $\Gamma=2.11\pm 0.12$ and normalization $A=4.1_{-1.1}^{+2.3}\times 10^{-12}$~(MeV cm$^2$ s)$^{-1}$.  The $\chi^2$ of the fit is $2.4$ for 7 degrees of freedom.   {The underlying} model is that of \gr\ emission from neutral pion decays produced by interactions of cosmic ray distribution which is powerlaw in momentum $p$: $dN_{CR}/dp\propto p^{-\Gamma_{CR}}$. Using the parametrization of pion decay spectrum from proton-proton interaction cross-sections from Ref. \cite{naima} one finds  $\Gamma_p=2.09_{-0.07}^{+0.06}$ from the \gr\ data fit in the energy range $200$~MeV$<E<400$~GeV. The $\chi^2$ of the fit is $10.5$ for 11 degrees of freedom.  

The spectrum of Con III is consistent with that of the Cygnus region cosmic ray cocoon in the Milky Way \cite{cygnus} also shown in Fig. \ref{fig:spectrum}. Hardness of the Cygnus cocoon spectrum has been interpreted as possibly being due to the presence of "fresh" cosmic rays injected from  Cygnus OB2  association \cite{cygnus} which is still too young to produce supernovae.  This implies that cosmic ray production could start before the onset of supernova activity in star forming regions.  However, superposition of the star forming complexes of different ages in Cygnus region \cite{orion_arm} precludes the possibility of firm association of cosmic ray population with parent stellar population. 

The projection problem is absent in the case of Con III region. Cosmic rays residing in the region are produced following the most recent star formation episode which occurred 10-15 Myr ago. Timing of the moment of the onset of cosmic ray production could be established from the spectral properties of the cosmic ray population. The spectrum of cosmic rays softens with time due to the energy-dependent diffusion of particles away from their production sites. Measurements in the Milky Way galaxy suggest that the diffusion coefficient scales with energy as $D(E)\simeq 3\times 10^{28}\left(E/10\mbox{ GeV}\right)^{\delta}$  \cite{blasi} with $\delta\simeq 0.33$ \cite{ams-02} so that the cosmic ray spectrum softens from the injection spectrum $dN/dE\propto E^{-\Gamma_0}$ to  $dN/dE\propto E^{-(\Gamma_0+\delta)}$ on the time scale  $t\sim R^2/D(E)\simeq 10\left(R/1\mbox{ kpc}\right)^2\left(E/10\mbox{ GeV}\right)^{-\delta}$~Myr of escape  from the production region of the size $R$ (about 1 kpc in Con III region, see Fig. \ref{fig:image}). 

If the  mechanical energy of massive star winds would provide a sizeable power for cosmic ray production in Con III region, injection of cosmic rays  would  have started 10-15 Myr ago and the spectrum of cosmic rays would be softened to $dN/dE\propto E^{-(\Gamma_0+\delta)}$ by now. Its slope would match the average slope of  cosmic ray spectrum of the Milky Way disk and of the LMC, $\overline \Gamma\simeq 2.4.... 2.5$ \cite{hard,yang,fermi_hard}.  To the contrary, the supernova rate in Con III region has reached a peak at the present epoch, see Fig. \ref{fig:sn_rate}.  Cosmic rays injected by supernovae still retain their injection spectrum. The slope of the cosmic ray spectrum in the region, $\Gamma_{CR}\simeq 2.1$ agrees with the  $\overline\Gamma-\delta=2.1...2.2$ injection spectrum slope inferred from the average Milky Way and LMC spectrum modelling \cite{hard,yang,fermi_hard}.  We conclude that the Con III data do not agree with the model of injection of cosmic rays before the onset of supernova activity and agree with the model of supernova origin of the cosmic rays. 

Supernova rate in the region is at the level of  ${\cal R}_{SN}\simeq 1/(2000$~yr)  since  ${\cal T}_{SN}\sim 5$~Myr (Fig. \ref{fig:sn_rate}). The cosmic ray energy ${\cal E}_{CR}$ injected by each supernova  is gradually transferred to gamma-rays on the time scale  of proton-proton interactions energy loss $t_{pp}=(c\kappa\sigma_{pp}n)^{-1} \simeq 3\times 10^8\left[n/0.3\mbox{ cm}^{-3}\right]^{-1}\mbox{ yr}$
where $\sigma_{pp}\simeq 3\times 10^{-26}\mbox{ cm}^2, \ \kappa\simeq 0.4$ are the cross-section and inelasticity of the proton/nuclei collisions \cite{aharonian}, $c$ is the speed of light  and $n$ is the density of interstellar medium. An estimate $n\simeq 0.3\left(H/500\mbox{ pc}\right)^{-1}$~cm$^{-3}$ could be derived from the measured column density atomic hydrogen $N_H\simeq 10^{21}$~cm$^{-2}$ at the position of Con III   \cite{hi} assuming thickness of the LMC disk $H\sim 500$~pc \cite{lmc_disk}.  The pion decay \gr\ luminosity produced by such energy release is $L_\gamma\simeq {\cal E}_{CR}/(3t_{pp})$ where   the factor $1/3$ takes into account the fact that only one third energy is deposited into neutral pions which decay into \gr s.  Cumulative flux of pion decay emission generated by cosmic rays ejected from some ${\cal T}_{SN}{\cal R}_{SN}\simeq (2.2_{-0.2}^{+1.0})\times 10^3$ supernovae accumulated in the Con III from the latest star formation episode is then  ${\cal F}_\gamma={\cal T}_{SN}{\cal R}_{SN}L_\gamma/\left(4\pi d_{LMC}^2\right)$ where $d_{LMC}\simeq 50$~kpc is the distance to the LMC \cite{lmc_disk}.
The energy output per supernova is 
\begin{equation}
{\cal E}_{SN}=\frac{12\pi d_{LMC}^2 {\cal F}_\gamma}{{\cal T}_{SN}{\cal R}_{SN}c\kappa\sigma_{pp} n}
\end{equation}
 Integrating the \gr\ flux over the energy range of Fermi/LAT data  one finds 
${\cal F}_\gamma= 3.5\pm 0.4\times 10^{-11}\mbox{ erg}/\mbox{(cm}^2\mbox{s)}$
which results in the estimate
\begin{equation}
{\cal E}_{SN}\simeq (1.1_{-0.2}^{+0.5})\times 10^{50}\left[\frac{n}{0.3\mbox{ cm}^{-3}}\right]^{-1}\mbox{ erg}
\end{equation}
where only uncertainties of the supernova statistics and of the \gr\ flux measurement  are taken into account. 

The \gr\ spectrum has a "bump" in the GeV range characteristic to the neutral pion decay \cite{snr1,tavani}. This shows that \gr\ emission is dominated by the pion decay component and that electron contribution to the \gr\ emission is small. This is consistent with the overall pattern of diffuse \gr\ emission from the Milky Way disk  where electrons contribute  10-20\% of the diffuse GeV \gr\ flux \cite{fermi_diffuse} and of  LMC outside 30 Dor region \cite{electrons}. 

{A limitation of analysis presented above is that is is based on observations of signal from a single star forming region. For each single source   measurement and model uncertainties (e.g. of the star formation history and of the model of diffusion of cosmic rays through the interstellar medium) might conspire in the estimate of the overall cosmic ray injection power and in conclusion about the dominance of the supernova contribution to the cosmic ray flux. It is important to observe other similar sources (isolated young star forming regions with controlled star formation history) to scrutinise the result. }  

{Another limitation  is that it refers to cosmic rays with energies up to TeV, while the Galactic cosmic ray spectrum presumably extends up to the PeV range. It is possible that the dominant source population changes with the increase of cosmic ray energy. Extension of analysis reported above into 1-100 TeV range (which will be possible with CTA)  is important in this respect.}

To summarize, timing of the moments of the onset of supernova activity and of cosmic ray production in Con III region has enabled identification of supernovae (rather than massive star wind driven superbubble) origin of cosmic rays. Calorimetric measurement of the cosmic ray content based on \gr\ signal combined with a measurement of the size of supernova population in Con III  has resulted in an estimate of  $\sim 10^{50}$~(erg/supernova)  cosmic ray injection energy, as expected in the supernova scenario of cosmic ray origin. \gr\ data also provide a measurement of the cosmic ray injection spectrum which is a powerlaw with the slope $\Gamma_{CR}\simeq 2.1$, as expected in the shock acceleration model. These two measurements{ are consistent with} the basic predictions of the  80-year-old conjecture of supernova origin of  cosmic rays  \cite{baade_zwicky,ginzburg} and of the shock acceleration model for cosmic ray production  \cite{fermi49,krymskii,bell1}, {in particular region of the LMC galaxy}.


\end{document}